# Role of transport performance on neuron cell morphology

E. Louis[1], C. Degli Esposti Boschi[2],
G.J. Ortega[3], E. Fernández[4]

[1]Departamento de Física Aplicada and Unidad Asociada of the Consejo Superior de Investigaciones Científicas, Universidad de Alicante, Apartado 99, E-03080 Alicante, Spain

[2]CNR-INFM, Unità di Ricerca di Bologna, viale Berti-Pichat, 6/2, I-40127, Bologna, Italy

[3]Departamento de Física, F.C.E.N. Universidad de Buenos Aires and CONICET, Pabellón I, Ciudad Universitaria, 1428, Capital Federal, Argentine

[4]Instituto de Bioingeniería, Universidad Miguel Hernández, Campus de San Juan, Apartado 18, E-03550 Alicante, Spain

**Address all correspondence and reprint requests to:**
Dr. Eduardo Fernández
Universidad Miguel Hernández
Department of Histology and Institute of Bioengineering
Fac. Medicina
San Juan 03550 ALICANTE
SPAIN
Tel.: +34/96 591 9439; FAX: +34/96 591 9434
Email: e.fernandez@umh.es

## Abstract

The compartmental model is a basic tool for studying signal propagation in neurons, and, if the model parameters are adequately defined, it can also be of help in the study of electrical or fluid transport. Here we show that the input resistance, in different networks which simulate the passive properties of neurons, is the result of an interplay between the relevant conductances, morphology and size. These results suggest that neurons must grow in such a way that facilitates the current flow. We propose that power consumption is an important factor by which neurons attain their final morphological appearance.

Manuscript Text

The actual morphology of neurons depends mainly on their own intrinsic properties and their connectivity. Their cell membranes are lipid bilayers with proteins floating around them, some of which are ion channels. The cytoplasm is composed mainly by water, electrolytes, charged proteins, cytoskeleton, endoplasmic reticulum and other organelles As a result, electrical signals propagate through the neuron cytoplasm that has a conductance $g_c$. Current can also flow through the membrane (having a conductance $g_m$) to the extracellular space [13]. The equivalent electrical structure is essentially a branching tree of axial resistors, mimicking the cytoplasm and connected together in a way that mirrors the morphology, (**Fig 1**). In addition a resistance associated to the output (synaptic resistance) has to be included [11, 13]. It is interesting to note that the values of those conductances may cover many orders of magnitude, depending on the system and on the environmental conditions [9].

Although the compartmental model has been used to calculate the passive electrical properties of a large variety of neurons [11, 13] a detailed comparison between the electrical performance of different cell morphologies is still lacking. This is so despite of the fact that such a study could throw light on the mechanisms that force neurons to branch. The present study is addressed to investigate how power consumption in different networks with largely different morphologies depend on the parameters of the compartmental model, namely cytoplasm, membrane and synaptic resistances. We examined whether patterns, mimicking the morphological appearance of neurons, grow in such a way that the transport efficiency is maximized (or, equivalently, minimizing the resistance to electrical transport).

Since neurons tend to ramify, creating large and complicated trees (see **Figs. 1a and 1b**), fractal systems can be useful to model their geometrical properties [6]. In our simulations we used Diffusion Limited Aggregation (DLA) networks (**Fig. 1c**) and a deterministic fractal, having fractal dimensions of 1.71 and 1.47, respectively [8, 14]. As it has been claimed that some types of neurons are rather space-filling objects [12], we also used the Eden model (**Figs. 1d**), which takes the main physics underlying the formation of many living systems [8] and represents a rapidly saturating system. We assumed that compartments with membrane conductance $g_m$, coupled through junctions of contact conductance $g_c$ (the cytoplasm conductance), were arranged in networks of coordination $p$ (**Fig. 1g**). The terminals were all assumed to be synaptic junctions with conductance $g_{syn}$. For the results presented here, both the extracellular space and the synaptic junctions were assumed to be isopotential. For simplicity reasons we did not take into account the thinning of neuron branches as the neuron grows.

If a current $I$ is injected at site $i$, which we assume to be located at the center of the network, Kirchoff's current law applied to the model outlined above, leads to a set of coupled linear equations for the potential at all sites $V_j$,

$$(1) \qquad I - g_m V_i = \sum_{j=1}^{n_i} g_c (V_i - V_j) + (p - n_i) g_{syn} V_i ,$$

where the first sum in the rhs of the above equation runs over the existing nearest-neighbors of site $i$, $n_i$, while the second accounts for its contact with the extracellular space and covers the remaining nearest-neighbors up to the actual coordination $p$ of the chosen lattice (**Fig. 1g**). The equation associated to a general element where no current is injected is exactly as in the above equation with $I=0$. Solving the resulting system of linear equations allows to obtain the input resistance of the aggregate $R_0=V_i/I$, where $V_i$ denotes the potential of the compartment at where current is injected. Hereafter we shall take $g_c$ as the unit of conductance. Therefore, including the outside medium and the synapses, our model represents a particular type of weighted network, where weights are given by the values of the relative conductances. Moreover, even though we are not pursuing an allometric study [15], we note that the above equations can be rewritten to describe fluid transport by replacing voltage by pressure and current intensity by fluid flow [3]. In solving the system of linear equations we have not used multigrid methods [4] nor algorithms based upon efficient suitable transformations [7]. Instead we have utilized the fast sparse matrix techniques available nowadays. These methods allowed us to handle compartments having up to $10^6$ sites.

The results of the numerical calculations illustrated in **Fig. 2** correspond to aggregates having $125^2$ sites. An extremely small value like $g_m=e^{-28}\,g_c$ has been included in order to show the effect on the low synaptic conductance tail. In the limit of high synaptic conductance, no current flows through the membrane and the input conductance depends only on the number of compartments the current should flow through, before reaching the synaptic junctions. As the synaptic conductance decreases the results indicate that networks can be classified in two groups: compact and ramified. Crossings between the curves for those two groups are observed, indicating that the optimal structure may vary depending on the relative values of the system conductances. At low synaptic conductance, ramified networks perform better (smaller input resistance) for the lower value of the membrane conductance (**Fig. 2a**), while the opposite occurs for $g_m=e^{-13}\,g_c$ (**Fig. 2b**). In the latter case the input resistance becomes constant at low values of the synaptic conductance, as current flows mainly through the cell membrane (this occurs for the square network even at the larger value of the membrane conductance as it has less synaptic connections, see **Fig. 2b**).

**Fig. 2c,d** shows the results of the input resistance $R_0$ versus the membrane conductance, for DLA and Eden aggregates of the same size as above, and output conductances of $g_m=e^{-17}\,g_c$ (**Fig. 2c**) and $g_m=e^{-5}\,g_c$ (**Fig. 2d**). A crossing of the results for the two types of aggregate is again observed for the lower value of the output conductance (**Fig. 2c**). When the output conductance is increased (**Fig. 2d**) no crossing occurs and the compact aggregate shows a lower input resistance in the whole range of membrane conductances shown in the Figure. At this stage, it is interesting to note that neurons, which usually have a low membrane conductance, show ramified structures [5, 10, 13], while, for instance, other networks of gap-junction connected cells such as the pancreatic β-cells, that have a much larger membrane conductance [1, 2] show compact structures. These experimental facts are in accordance with the results of **Fig. 2c**, namely, while at very low membrane conductances the DLA aggregate shows a smaller input resistance, the opposite occurs at intermediate values of $g_m$. For very high membrane conductances the whole current flows through the axial resistance and all networks show an input resistance that decreases as $g_m$ increases.

We have also investigated how these results depend on the system size. In **Fig. 3** we show the results of the input resistance $R_0$ versus the number of compartments in the network, $N$, for systems with a very small membrane conductance $g_m=e^{-17}g_c$ and a synaptic conductance either 20 times smaller (**Fig. 3a**) or equal to the contact conductance (**Fig. 3b**). In the first case $R_0$ decreases approximately as $1/N$ no matter which system morphology is considered. This is readily understood by noting that in such a case the system behaves as a network of $N$ resistances $g_m^{-1}$ in parallel as all compartments are approximately at the same voltage. When $g_{syn} = g_c$ no current flows through the cell membrane and all is carried to the extracellular space through synaptic junctions. As a consequence, compact systems show an input resistance that increases logarithmically with $N$, a behavior typical of two dimensional systems, while in ramified networks $R_0$ becomes constant beyond a value of $N$ that depends on the actual ratio $g_m/g_{syn}$. This is further illustrated by the results for intermediate values of $g_{syn}$. As shown in **Fig. 3c,d**, the input resistance undergoes a crossover from a power law to either a constant $R_0$ in ramified systems, or, as shown in **Fig. 3b**, a logarithmic behavior in the case of compact systems. It is interesting to note that, while the power law followed by ramified systems at sufficiently small $N$ is always close to $1/N$ (a sign of good scaling down to small sizes), the power steadily decreases with $g_{syn}$ in compact systems. The crossover occurs at a characteristic size that depends on the actual values of the conductance and crucially on the network type (whether compact or ramified). An important consequence is that, while at small $N$ compact networks show a higher resistance and, thus, ramified networks would be more efficient in carrying currents, the trend changes in favor of the compact ones when $N$ increases. This indicates that a crossover from ramified to compact may occur as the system grows.

Our results suggest that the input resistance of a network, which quantifies the passive response of a neuron to a current feed, determines to some extent its growth morphology. We believe that although the model investigated here is for sure an oversimplification of the complex morphology of neurons, it may help to understand how the dendritic morphology influences the input-output function of a neuron and shed light on a problem to which little attention has been drawn: the dependence of transport performance on neuron morphology.

## ACKOWLEDGMENTS

This work was supported by the Spanish Comisión Interministerial de Ciencia y Tecnología" through grants PB96 0085, 1FD97 1358, TIC2003-09557-CO2-02 and by the European Commission through the projects "TMR Network-Fractals c.n. FMRXCT980183", "FET Open project COSIN IST-2001-33555" and "QLK6-CT 2001-00279". CDEB acknowledges a postdoctoral position at the Universidad de Alicante, during which this work was started.

# Figures

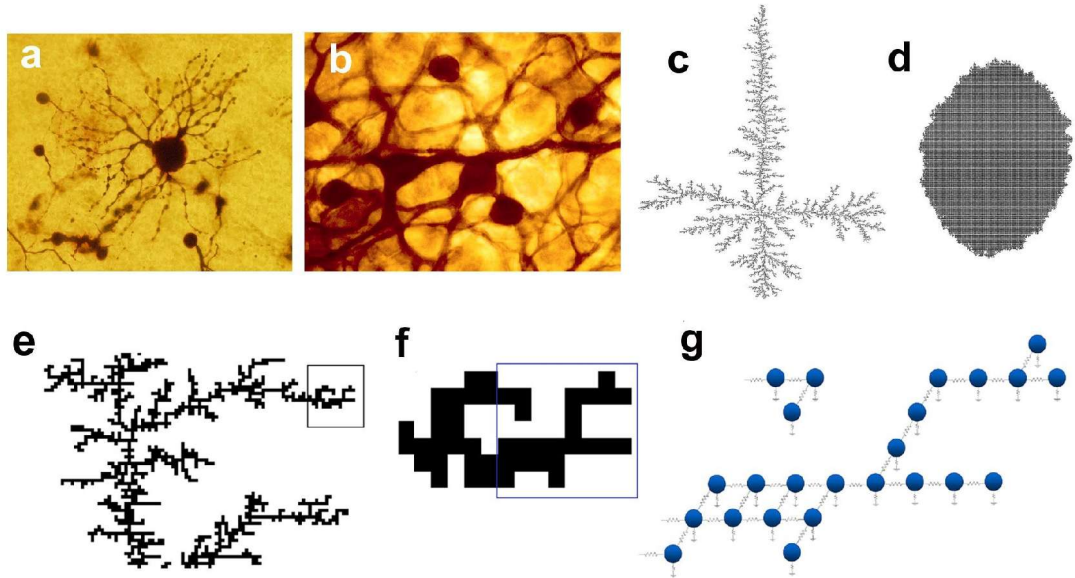

**Figure 1.** Examples of neurons, simulated networks and the simplified electrical representation of a segment of passive cable. (**a**) Golgi-impregnation of a human retinal ganglion cell. (**b**) Neurobiotin labeling of horizontal cells in turtle retina. (**c**) Typical diffusion-limited aggregation (DLA) fractal. (**d**) Compact network generated by the Eden model. (**e**) Detail of c. (**f**) High magnification of the insert in e. (**g**) Compartmental electrical representation of the DLA segment showed in f. Several membrane patches are connected in series via axial resistances. The extracellular space is assumed to be isopotential. This is a discrete modeling of a continuous cable.

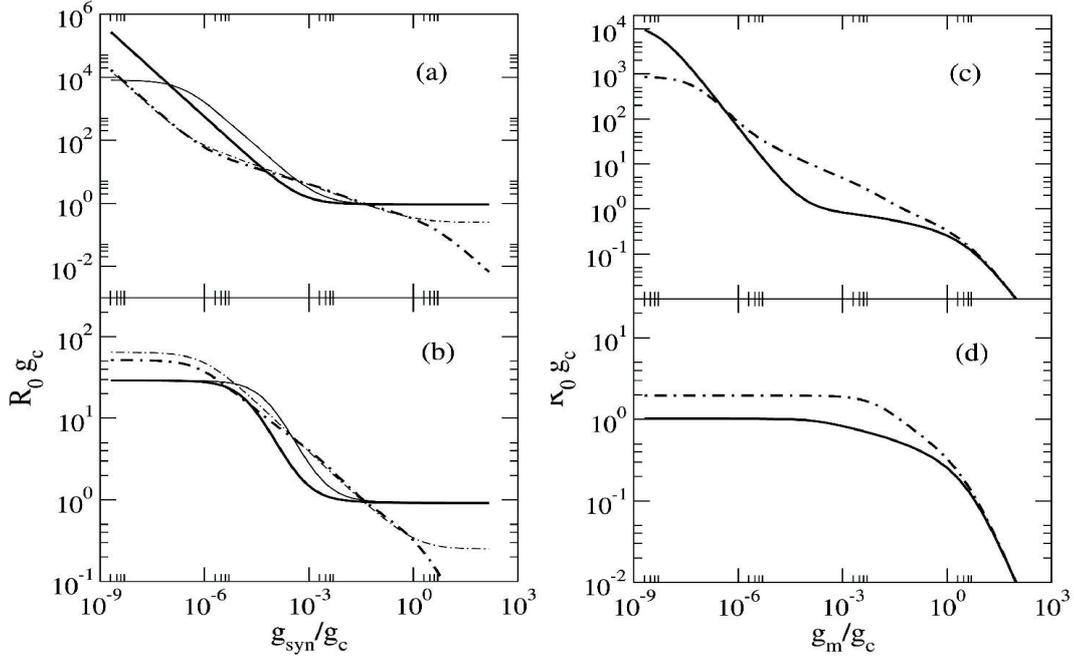

**Figure 2.** Input resistance $R_0$ of systems with $N=125^2$ compartments versus the conductance that couples the network to the extracellular space $g_{syn}$ (a,b) and the membrane conductance $g_m$ (c,d). In (a) and (b) the membrane conductance was fixed at $g_m=e^{-28} g_c$ and $g_m=e^{-13} g_c$, respectively, while in (c) and (d) the synaptic conductance was kept constant at $g_{syn}=e^{-17} g_c$ and $g_{syn}=e^{-5} g_c$ (**d**). The symbols correspond to DLA aggregates (thick chain line), a deterministic fractal (thin chain line), a perfect square (thin continuous line) and the Eden model (thick continuous line).

**Figure 3.** Numerical results for the input resistance $R_0$ (in units of $g_c^{-1}$) versus the number of compartments in the system $N$. The symbols correspond to: DLA (diamonds), deterministic fractal (crosses), a perfect square (squares) and the Eden model (circles).

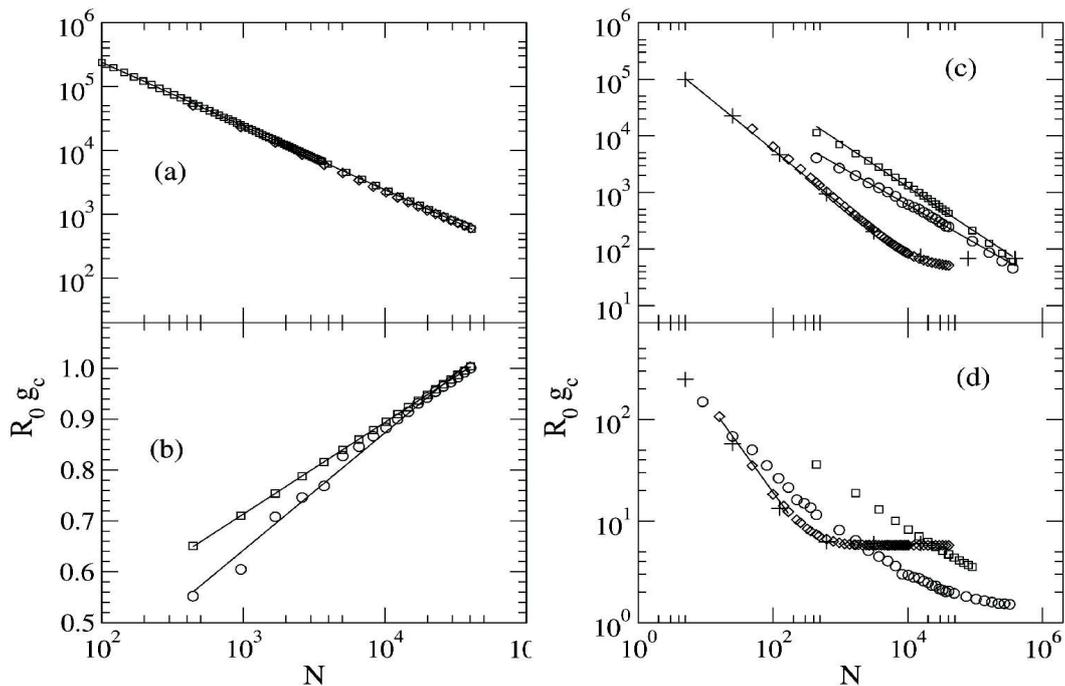

(a) $g_m = e^{-17} g_c$ and $g_{syn} = e^{-20} g_c$. All results collapse (only some of them are shown) on the curve $R_0 = 2.4 \times 10^7 \, N^{-0.98} \, g_c^{-1}$. (b) $g_m = e^{-17} g_c$ and $g_{syn} = g_c$. The curves fitted to the numerical results are $R_0 = (0.078 \ln N + 0.17) \, g_c^{-1}$ (perfect square) and $R_0 = (0.1 \ln N - 0.05) \, g_c^{-1}$ (Eden model). For ramified systems (both DLA and deterministic) $R_0$ readily become constant (0.32 and 0.34 respectively). (c) $g_m = e^{-17} g_c$ and $g_{syn} = e^{-14} g_c$. The square and the Eden model follow a power law in the whole range with fittings $R_0 = 1.8 \times 10^6 \, N^{-0.79} \, g_c^{-1}$ and $R_0 = 3.0 \times 10^5 \, N^{-0.67} \, g_c^{-1}$ respectively whereas $R_0$ for ramified systems (DLA or deterministic fractal) is constant beyond $N \cong 10^4$ and follow the power law $R_0 = 4.9 \times 10^5 \, N^{-0.97} \, g_c^{-1}$ for smaller systems. (d) $g_m = e^{-17} g_c$ and $g_{syn} = e^{-8} g_c$. The results for ramified systems (DLA and deterministic) follow the power law $R_0 = 1.34 \times 10^3 \, N^{-0.92} \, g_c^{-1}$ for $N$ smaller than $10^3$, and remain constant for larger systems.